\documentclass[conference]{IEEEtran}
\IEEEoverridecommandlockouts
\usepackage{cite}
\usepackage{amsmath,amssymb,amsfonts}
\usepackage{algorithmic}
\usepackage{graphicx}
\usepackage{textcomp}
\usepackage{xcolor}
\usepackage{multirow}
\def\BibTeX{{\rm B\kern-.05em{\sc i\kern-.025em b}\kern-.08em
    T\kern-.1667em\lower.7ex\hbox{E}\kern-.125emX}}
\begin{document}

\title{Rapid Adaptation of SpO$_2$ Estimation to Wearable Devices via Transfer Learning on Low-Sampling-Rate PPG\\

}
\author{
\IEEEauthorblockN{
Zequan Liang$^{1}$,
Ruoyu Zhang$^{2}$,
Wei Shao$^{1}$,
Krishna Karthik$^{2}$,\\
Ehsan Kourkchi$^{2}$,
Setareh Rafatirad$^{1}$,
Houman Homayoun$^{2}$
} \\
\IEEEauthorblockA{
$^{1}$Department of Computer Science, University of California, Davis, Davis, CA, U.S.A.\\ 
$^{2}$Department of Electrical and Computer Engineering, University of California, Davis, Davis, CA, U.S.A.\\ 
Email: \{zqliang, ryuzhang, wayshao, kbhaskar, ekay, srafatirad, hhomayoun\}@ucdavis.edu}}

% \author{\IEEEauthorblockN{ Ruoyu Zhang}
% \IEEEauthorblockA{
% \textit{University of California, Davis}\\
% Davis, USA \\
% ryuzhang@ucdavis.edu}
% \and
% \IEEEauthorblockN{Ruijie Fang}
% \IEEEauthorblockA{
% \textit{ University of California, Davis}\\
% Davis, USA \\
% rjfang@ucdavis.edu}
% \and
% \IEEEauthorblockN{Elahe Hosseini}
% \IEEEauthorblockA{
% \textit{University of California, Davis}\\
% Davis, USA \\
% ehosseini@ucdavis.edu}
% \\
% \and

% \IEEEauthorblockN{Chongzhou Fang}
% \IEEEauthorblockA{
% \textit{University of California, Davis}\\
% Davis, USA \\
% czfang@ucdavis.edu}

% \and

% \IEEEauthorblockN{Ning Miao}
% \IEEEauthorblockA{
% \textit{University of California, Davis}\\
% Davis, USA \\
% nmiao@ucdavis.edu}
% \and
% \IEEEauthorblockN{Houman Homayoun}
% \IEEEauthorblockA{
% \textit{University of California, Davis}\\
% Davis, USA \\
% hhomayoun@ucdavis.edu}
% }

\maketitle

\begin{abstract}

Blood oxygen saturation (SpO$_2$) is a vital marker for healthcare monitoring. Traditional SpO$_2$ estimation methods often rely on complex clinical calibration, making them unsuitable for low-power, wearable applications. In this paper, we propose a transfer learning–based framework for the rapid adaptation of SpO$_2$ estimation to energy-efficient wearable devices using low-sampling-rate (25Hz) dual-channel photoplethysmography (PPG). We first pretrain a bidirectional Long Short-Term Memory (BiLSTM) model with self-attention on a public clinical dataset, then fine-tune it using data collected from our wearable We-Be band and an FDA-approved reference pulse oximeter. Experimental results show that our approach achieves a mean absolute error (MAE) of 2.967\% on the public dataset and 2.624\% on the private dataset, significantly outperforming traditional calibration and non-transferred machine learning baselines. Moreover, using 25Hz PPG reduces power consumption by 40\% compared to 100Hz, excluding baseline draw. Our method also attains an MAE of 3.284\% in instantaneous SpO$_2$ prediction, effectively capturing rapid fluctuations. These results demonstrate the rapid adaptation of accurate, low-power SpO$_2$ monitoring on wearable devices without the need for clinical calibration.
\end{abstract}


\section{Introduction}

Blood oxygen saturation (SpO$_2$) is a critical physiological parameter that reflects the percentage of oxygen-bound hemoglobin in the blood, providing essential insights into cardiovascular function \cite{evans2001vital,fang2022pain}. Pulse oximetry estimates SpO$_2$ non-invasively using photoplethysmography (PPG) sensors by analyzing the light absorption of pulsatile blood flow, typically red and infrared (IR) wavelengths \cite{kumar2021pulse}.

\begin{figure}[ht]
    \centering
    \includegraphics[width= \linewidth]{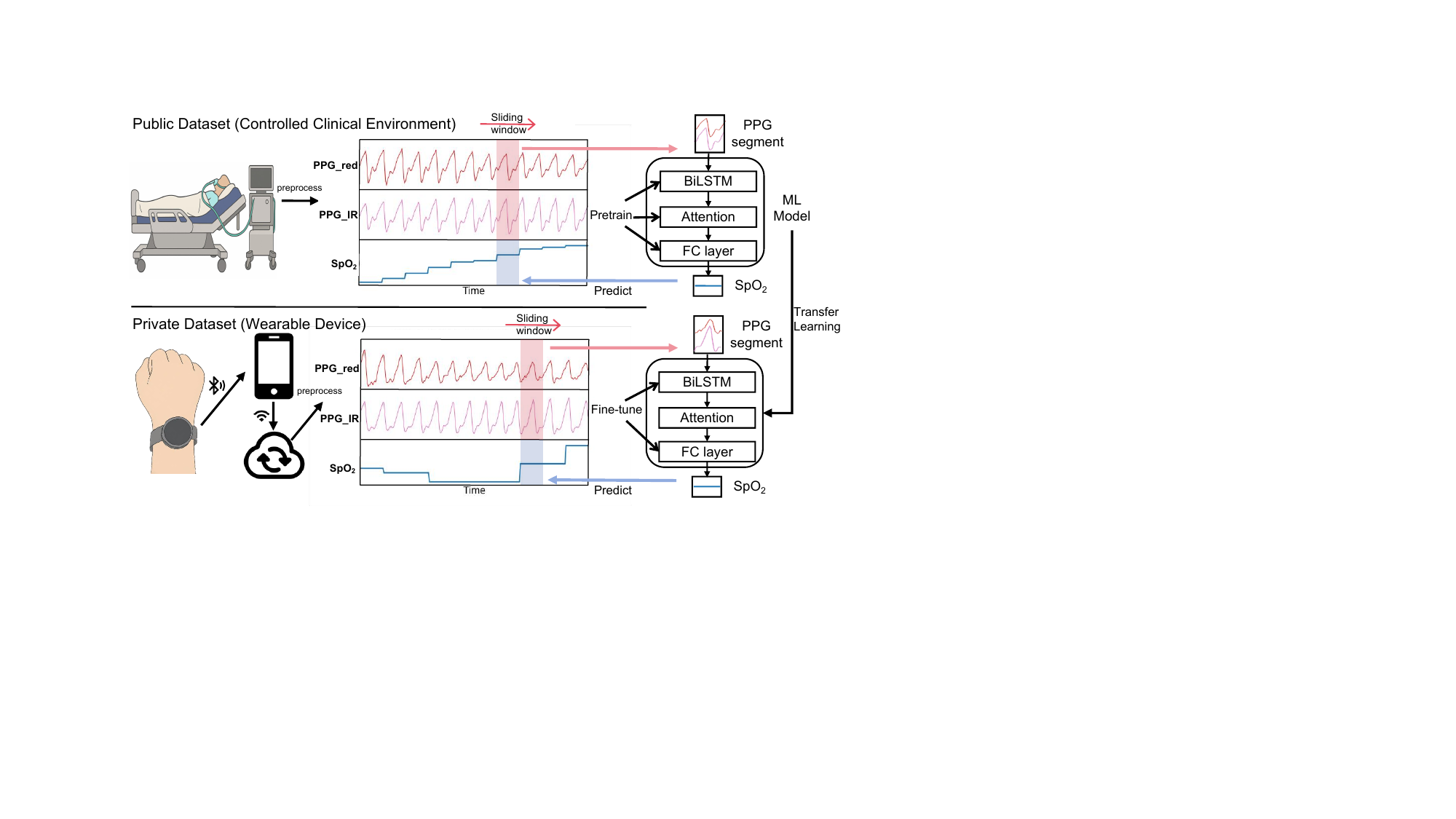}
    \caption{SpO$_2$ estimation framework using transfer learning}
    \label{fig:big_picture}
\end{figure}

In recent years, machine learning models have been employed for SpO$_2$ estimation from PPG signals. 
% \cite{shuzan2023machine,venkat2019machine,koteska2022deep}. 
However, several challenges remain for practical deployment in wearable health monitoring systems. First, many existing approaches rely solely on single-channel input \cite{shuzan2023machine} and require high PPG sampling rates, which leads to increased power consumption. Second, traditional SpO$_2$ calibration on new devices often requires controlled clinical calibration, which is impractical for rapid deployment \cite{fong2025open}. Lastly, spontaneous SpO$_2$ fluctuations are poorly represented in clinical datasets where saturation levels are generally maintained within stable ranges \cite{fong2025open}.

To address these limitations, we propose a transfer learning–based framework for rapid adaptation of SpO$_2$ estimation to new wearable devices using low-rate dual-channel PPG. As shown in Fig.~\ref{fig:big_picture}, our approach mainly pretrains a bidirectional Long Short-Term Memory (BiLSTM) model with self-attention \cite{li2020bidirectional} on a publicly available dataset collected in a controlled clinical environment, and then fine-tunes it using a private dataset collected from the wearable We-Be band \cite{zhang2024introducing}. 

This work makes the following key contributions: \textbf{(1) Low-Sampling-Rate PPG Enables Energy-Efficient Wearable Devices:} Our system operates at a sampling rate of 25Hz, reducing power consumption and extending the battery life of wearable devices for long-term health monitoring. \textbf{2) Transfer Learning Enables Rapid Deployment Across Devices:} We enable fast model adaptation on a wearable device through transfer learning by fine-tuning on small amounts of user-level data, eliminating the need for clinical calibration. \textbf{3) Capturing Instantaneous SpO$_2$ Fluctuations:} Our model is capable of detecting rapid SpO$_2$ changes, which are essential for early warning in conditions such as sleep apnea and respiratory distress.

% This approach makes the following key contributions: 1) Low-Sampling-Rate PPG Enables Energy-Efficient Wearable Devices:
% Our system operates at a sampling rate of 25Hz, resulting in a 17.6\% reduction in power consumption compared to the commonly used 128Hz, thereby extending battery life for long-term monitoring. Despite the lower sampling rate, the model achieves a mean absolute error (MAE) of 2.967\% on the public dataset, demonstrating that SpO$_2$ information can still be effectively extracted. 2) Transfer Learning Enables Rapid Deployment Across Devices: By fine-tuning on a small amount of user-collected data from a wearable device, our model achieves an MAE of 3.373\% on the private dataset, outperforming traditional calibration-based methods. This demonstrates that accurate SpO$_2$ estimation can be rapidly deployed without requiring controlled clinical calibration. 3) Capturing Instantaneous SpO$_2$ Fluctuations: Our method achieves an MAE of 3.629\% in instantaneous SpO$_2$ prediction, showing its ability to detect rapid changes in SpO$_2$, which are essential for early warning in conditions such as sleep apnea and respiratory distress.

\section{Framework Overview} \label{Working Principle}
The overall framework is illustrated in Fig.~\ref{fig:big_picture}. We utilize the OpenOximetry Dataset \cite{fong2025open}, a publicly available dataset collected under a medical-grade clinical environment. Red (660nm) and infrared (940nm) PPG signals were recorded at a sampling rate of 86Hz, along with reference SpO$_2$ values obtained by averaging readings from multiple pulse oximeters. To align with our We-Be band \cite{fang2024validation}, which operates at 25Hz, we downsample the PPG signals from 86Hz to 25Hz by Fourier-domain resampling. After preprocessing, both red and IR signals are then segmented using a sliding window. Each resulting segment is paired with its corresponding SpO$_2$ value, enabling supervised learning. BiLSTM has been widely adopted for temporal feature extraction \cite{liang2025multi}. Our machine learning (ML) model includes a BiLSTM, a self-attention layer, and a fully connected (FC) layer. They are pretrained jointly to learn PPG patterns and regress the SpO$_2$ values.

\begin{figure}[htbp!]
    \centering
    \includegraphics[width=0.45\linewidth]{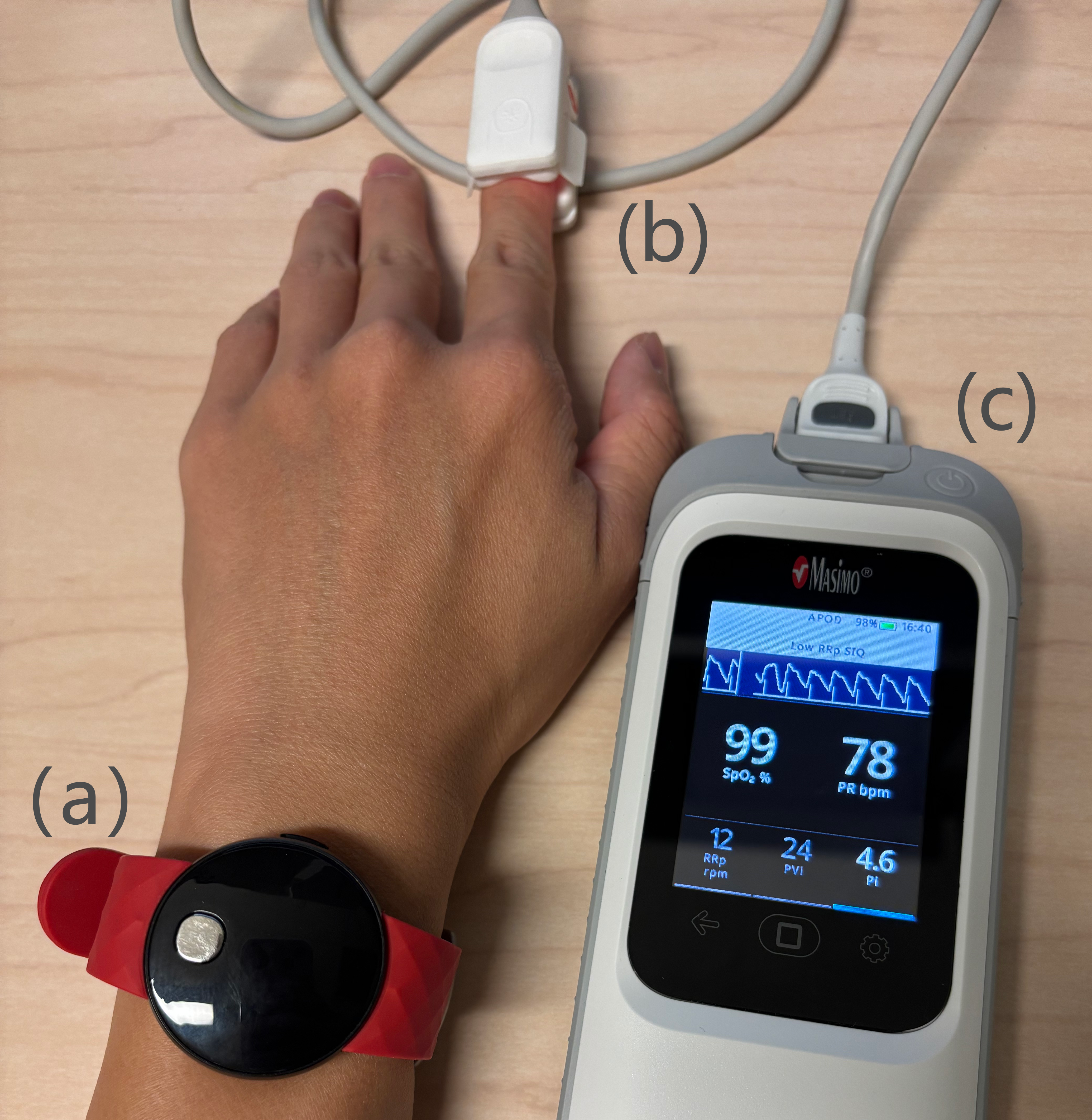}
    \caption{Private wearable data collection. (a) We-Be band for PPG signals (b) Fingertip sensor attached to Masimo Rad-G (c) Masimo Rad-G  for reference SpO$_2$ }
    \label{fig:SPO2_collection}
\end{figure}

To make the rapid adaptation, we collect a private dataset, as illustrated in Fig.~\ref{fig:SPO2_collection}, using the We-Be band equipped with PPG sensors (660nm red and 950nm IR, sampled at 25Hz), alongside an FDA-approved Masimo Rad-G pulse oximeter as the reference. To induce transient SpO$_2$ fluctuations, participants perform at least 3 cycles of breath-holding. 
% In our system, data is synchronized via a smartphone application using Bluetooth and stored on a cloud server. 
We apply the same preprocessing and segmentation procedures to the private dataset and fine-tune the pretrained ML model. Specifically, we fine-tune the BiLSTM and the FC layer to regress the reference SpO$_2$ values better. 

After transfer learning, the resulting model is deployed on the We-Be band for real-time SpO$_2$ estimation. It continuously processes dual-channel PPG segments and outputs a continuous stream of SpO$_2$ predictions.

\begin{figure}[t]
    \centering
    \includegraphics[width= \linewidth]{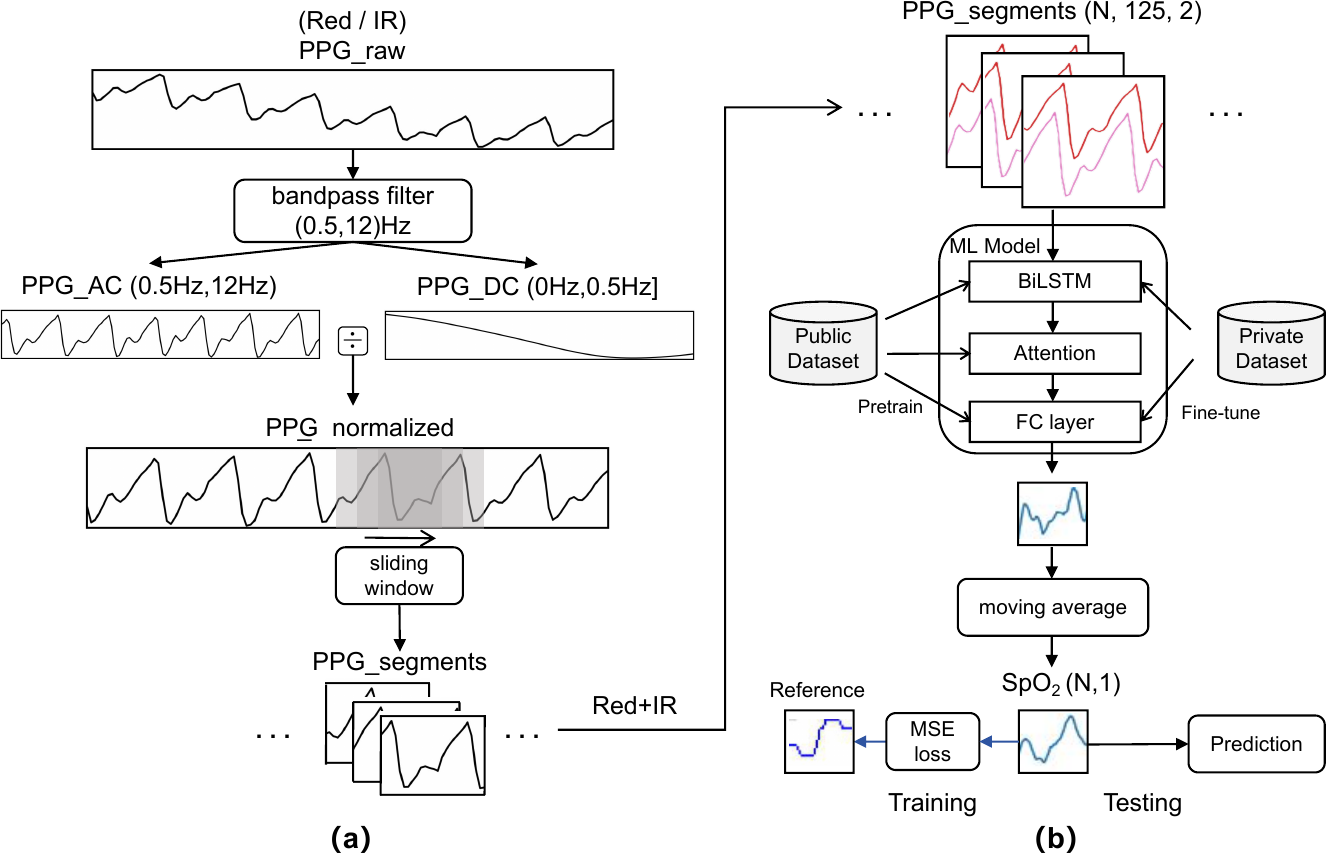}
    \caption{Data processing (a) PPG preprocessing (b) Machine learning method}
    \label{fig:SPO2_process}
\end{figure}

\section{Data Preprocessing and Model Architecture} \label{Prototype Development}

The data preprocessing pipeline is illustrated in Fig.~\ref{fig:SPO2_process}(a). We begin by extracting the raw PPG signals from both red and infrared (IR) sensors. To remove baseline drift and high-frequency noise, we apply a band-pass filter with a frequency range of 0.5–12Hz. This filter separates the signal into its alternating current (AC) and direct current (DC) components: the DC component captures low-frequency baseline trends below 0.5Hz, while the AC component captures the pulsatile variation of blood flow within the 0.5–12Hz range. To normalize the PPG signal, we compute the ratio of the AC to DC components. Next, we segment the normalized signal using a sliding window of 5 seconds with a 1-second stride. Given the 25Hz sampling rate, each window contains 125 sample points. 

% Over a full recording session, this yields $N$ overlapping PPG segments. 

As shown in Fig.~\ref{fig:SPO2_process}(b), we concatenate the red and IR PPG segments along the last dimension, resulting in a segment of shape $(125, 2)$. Across all $N$ segments, the input tensor to the model has shape $(N, 125, 2)$. Our ML model consists of three main components: a BiLSTM to learn temporal dependencies in the dual-channel PPG sequence, a self-attention layer that uses the internal BiLSTM features as the query, key, and value \cite{li2020bidirectional} to capture the global contextual relationship, and an FC layer to regress the final SpO$_2$ value. The model outputs a single SpO$_2$ prediction per window via the FC layer. A moving average filter with a window size of 5 is applied to the series of SpO$_2$ predictions, resulting in a final SpO$_2$ vector of shape $(N, 1)$. During training, we optimize the model using the mean squared error (MSE) loss between predicted and reference SpO$_2$ values. During testing, the output is used directly as the SpO$_2$ prediction. 

We first pretrain all model layers on the public dataset collected under controlled clinical conditions. Then we use our private wearable dataset to fine-tune only the FC layer. Finally, we unfreeze the BiLSTM and continue training both the BiLSTM and the FC layer together. This two-stage fine-tuning strategy results in a robust and low-cost solution for SpO$_2$ estimation on wearable platforms. 

\section{Experiment on Public Dataset}

To pretrain the model for SpO$_2$ estimation, we first conducted experiments on the public OpenOximetry dataset. The dataset was split at the subject level into training and test sets using a 4:1 ratio. Additionally, 5-fold cross-validation was performed within the training set. To match the characteristics of our We-Be band, we downsampled the original PPG signals from 86 Hz to 25 Hz by Fourier-domain resampling after a 0.5-12Hz band-pass filter. All models were implemented in PyTorch and trained using the Adam optimizer with a learning rate of 0.001. The batch size was set to 256, and each model was trained for 100 epochs using the MSE loss function. To address label imbalance in SpO$_2$, we also apply weighted sampling during fine-tuning. The SpO$_2$ labels were discretized into 10 bins, and sampling weights were assigned as the inverse bin frequencies. 

% To reduce prediction noise, we applied a moving average filter with a window size of 5 across the predicted SpO$_2$ sequence.

We compared the following five methods:
\begin{itemize}
\item Traditional\cite{guo2015reflective}: A traditional method that computes the $R$ ratio for quadratic SpO$_2$ calibration:

\begin{equation}
R = \frac{AC_{\mathrm{RED}} / DC_{\mathrm{RED}}}{AC_{\mathrm{IR}} / DC_{\mathrm{IR}}}
\end{equation}

\item CNN\_1d: A one-dimensional convolutional neural network that applies temporal convolution layers followed by a fully connected layer for regression.

\item Transformer: A transformer encoder with 4 self-attention heads applied to segmented PPG for sequence modeling.

\item BiLSTM: A two-layer bidirectional Long Short-Term Memory network that captures temporal dependencies in the PPG signal.

\item BiLSTM+Attn: Based on BiLSTM, this model integrates a self-attention mechanism to emphasize internal features in the temporal representation. 

\end{itemize}

\begin{table}[t]
\caption{SpO$_2$ Prediction Errors On Public OpenOximetry Dataset}
\begin{center}
\begin{tabular}{|c|c|c|c|c|}
\hline
\multirow{2}{*}{\textbf{Method}} & \multicolumn{2}{c|}{\textbf{25Hz}} & \multicolumn{2}{c|}{\textbf{86Hz}} \\
\cline{2-5}
& \textbf{\textit{MAE}} & \textbf{\textit{RMSE}} & \textbf{\textit{MAE}} & \textbf{\textit{RMSE}} \\
\hline
Traditional     & 3.286 & 4.892 & 3.280 & 4.880 \\
\hline
CNN\_1d         & 5.899 & 7.047 & 5.687 & 6.997 \\
\hline
Transformer     & 3.577 & 5.151 & 4.208 & 6.019 \\
\hline
BiLSTM          & 3.216 & 4.787 & 3.049 & 4.551 \\
\hline
BiLSTM+Attn     & \textbf{2.967} & \textbf{4.393} & \textbf{2.910} & \textbf{4.331} \\
\hline
\end{tabular}
\label{tab:public_spo2}
\end{center}
\end{table}

% \begin{table}[htbp]
% \caption{SpO$_2$ Prediction Errors on Public OpenOximetry Dataset}
% \begin{center}
% \begin{tabular}{|l|cc|cc|}
% \hline
% \textbf{Method} 
%     & \multicolumn{2}{c|}{\textbf{25Hz}} 
%     & \multicolumn{2}{c|}{\textbf{86Hz}} \\
% \cline{2-5}
%     & \textbf{MAE} & \textbf{RMSE} 
%     & \textbf{MAE} & \textbf{RMSE} \\
% \hline
% Traditional     & 3.286 & 4.892    & 3.280 & 4.880    \\
% CNN\_1d         & 5.899 & 7.047    & 5.687 & 6.997    \\
% TCN             & 3.112 & 4.446    & 3.332 & 4.769    \\
% BiLSTM          & 3.216 & 4.787    & 3.049 & 4.551    \\
% BiLSTM+Attn     & \textbf{2.967} & \textbf{4.393} 
%                 & \textbf{2.910} & \textbf{4.331} \\
% Transformer     & 3.577 & 5.151    & 4.208 & 6.019    \\
% \hline
% \end{tabular}
% \label{tab:public_spo2}
% \end{center}
% \end{table}

\begin{figure}[b]
    \centering
    \includegraphics[width= \linewidth]{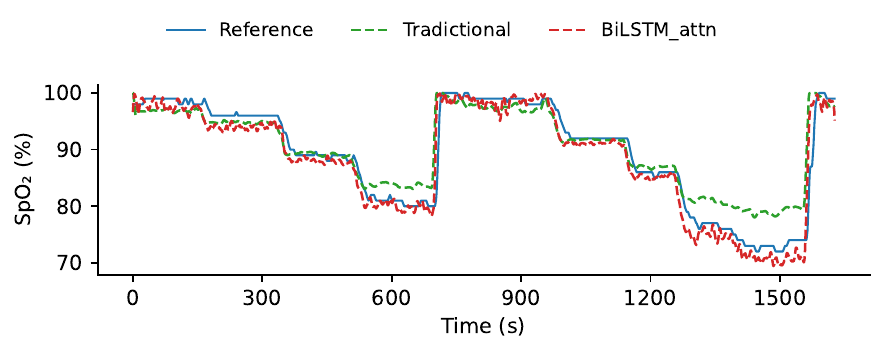}
    \caption{SpO$_2$ prediction results on a representative test case from public OpenOximetry Dataset using 25Hz PPG}
    \label{fig:public_spo2}
\end{figure}

Table~\ref{tab:public_spo2} presents the performance of all methods on the public OpenOximetry dataset in terms of Mean Absolute Error (MAE) and Root Mean Square Error (RMSE). Notably, the prediction error at 25Hz remains close to that before downsampling from 86Hz, indicating that key SpO$_2$ information is largely preserved. Even after downsampling to 25Hz, the BiLSTM+Attn model maintains the best overall performance, with an MAE of 2.967\% and an RMSE of 4.393\%. Fig.~\ref{fig:public_spo2} illustrates a representative test case using 25Hz PPG. Compared to the reference SpO$_2$ (blue solid line), the traditional method (green dashed) fails to accurately track low SpO$_2$ levels below 85\%. In contrast, the BiLSTM+Attn model (red dashed) closely follows the reference, demonstrating its effectiveness in capturing complex signal dynamics.

% These results confirm that the BiLSTM+Attn model can effectively extract SpO$_2$ information even from dual-channel PPG signals downsampled at a low rate of 25Hz, and outperform traditional calibration-based methods on the public OpenOximetry dataset.

\section{Experiment on Private Dataset}

\begin{figure}[t]
    \centering
    \includegraphics[width= 0.9\linewidth]{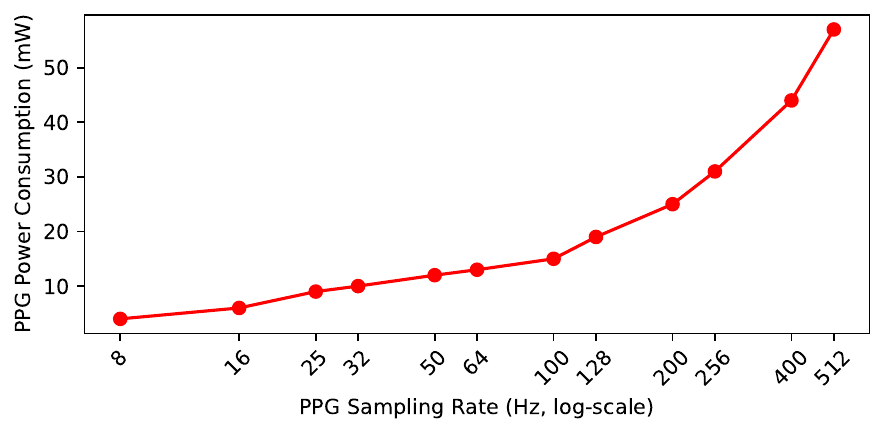}
    \caption{Power consumption of the PPG sensor on the We-Be band}
    \label{fig:sample_rate}
\end{figure}

\begin{figure}[t]
    \centering
    \includegraphics[width=\linewidth,height=0.12\textheight]{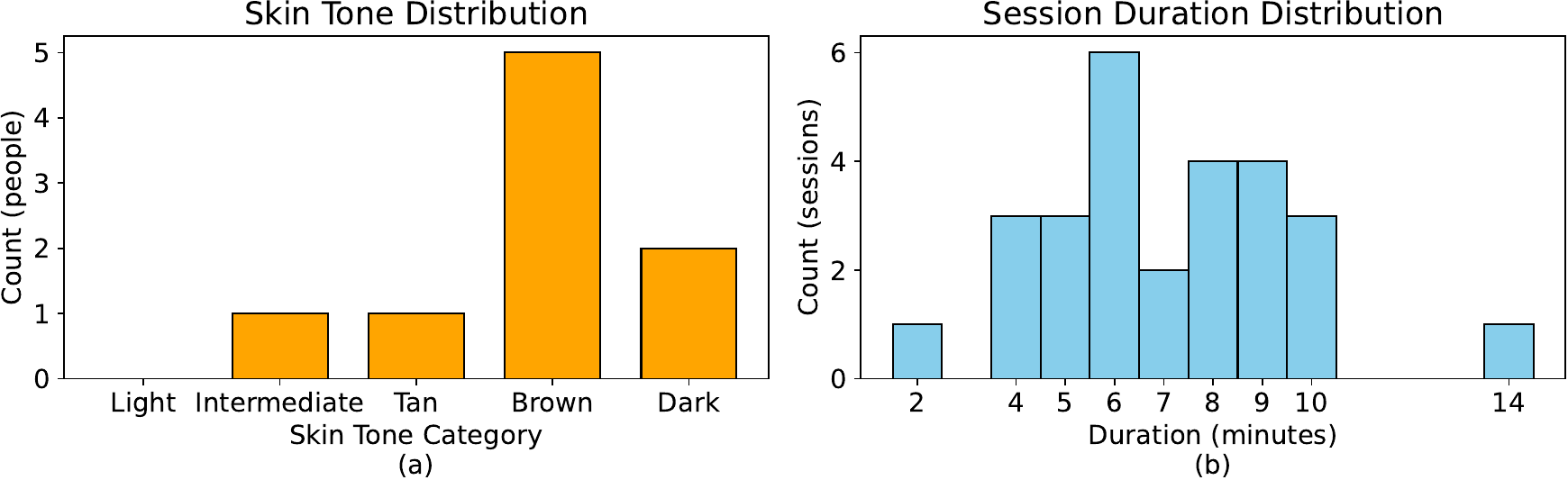}
    \caption{Distributions of the private dataset (a) Skin tones (b) Session durations}
    \label{fig:skin_tone_session_distribution}
\end{figure}

As shown in Fig.~\ref{fig:sample_rate}, we measured the power consumption of the PPG sensor across various sampling rates on the We-Be band. After subtracting the static power drawn by non-PPG components, the We-Be band consumes only around 9mW at 25Hz PPG, which is 40\% lower than around 15mW required at the commonly used 100Hz PPG. This demonstrates the significant energy savings achieved by using a lower sampling frequency.

To evaluate the effectiveness of our model in non-clinical conditions, we conducted experiments on a private dataset collected using wearable devices. The dataset includes dual-channel 25Hz PPG signals from We-Be band, along with corresponding SpO$_2$ reference values from Masimo Rad-G. In total, we collected 27 sessions from 9 individuals, with 3 sessions per person. We also measured wrist skin color using a Vinckolor colorimeter to classify their skin tones \cite{ly2020research}. Fig.~\ref{fig:skin_tone_session_distribution} summarizes the distributions of skin tone categories and session durations. Due to the limited sample size, we employed a leave-one-subject-out (LOSO) cross-validation strategy to ensure subject-independent evaluation.

We compared the following methods as before:
\begin{itemize}
\item Traditional: The same calibration method as above.

% \item BiLSTM: A pretrained BiLSTM model tested directly on the private dataset without fine-tuning.

% \item BiLSTM+Transfer: A pretrained BiLSTM model is fine-tuned on the private dataset during transfer learning.

\item BiLSTM+Attn: The pretrained BiLSTM model with a self-attention layer, tested directly on the private dataset.

\item BiLSTM+Attn+Transfer: A transfer learning approach where the pretrained BiLSTM with self-attention is fine-tuned on the private dataset for 150 epochs.
\end{itemize}

% \begin{table}[t]
% \caption{SpO$_2$ Prediction Errors On Private Wearable Dataset Using 25Hz PPG}
% \begin{center}
% \begin{tabular}{|c|c|c|c|c|}
% \hline
% \textbf{Method} & \textbf{MAE} & \textbf{RMSE} & \textbf{MAE$_{\text{ins}}$} & \textbf{RMSE$_{\text{ins}}$} \\
% \hline
% Traditional          & 4.069 & 5.228 & 4.738 & 5.780 \\
% \hline
% BiLSTM               & 2.696 & 3.566 & 4.242 & 5.285 \\
% \hline
% BiLSTM+Transfer         & \textbf{2.368} & \textbf{2.949} & \textbf{3.107} & \textbf{3.826} \\
% \hline
% BiLSTM+Attn          & 2.717 & 3.717 & 4.635 & 5.734 \\
% \hline
% BiLSTM+Attn+Transfer    & 2.624 & 3.214 & 3.284 & 3.999 \\
% \hline
% \end{tabular}
% \label{tab:webe_spo2}
% \end{center}
% \end{table}

\begin{table}[t]
\caption{SpO$_2$ Prediction Errors On Private Wearable Dataset Using 25Hz PPG}
\begin{center}
\begin{tabular}{|c|c|c|c|c|}
\hline
\textbf{Method} & \textbf{MAE} & \textbf{RMSE} & \textbf{MAE$_{\text{ins}}$} & \textbf{RMSE$_{\text{ins}}$} \\
\hline
Traditional          & 4.069 & 5.228 & 4.738 & 5.780 \\
\hline
BiLSTM+Attn          & 2.717 & 3.717 & 4.635 & 5.734 \\
\hline
BiLSTM+Attn+Transfer & \textbf{2.624} & \textbf{3.214} & \textbf{3.284} & \textbf{3.999} \\
\hline
\end{tabular}
\label{tab:webe_spo2}
\end{center}
\end{table}

\begin{figure}[t]
    \centering
    \includegraphics[width=\linewidth]{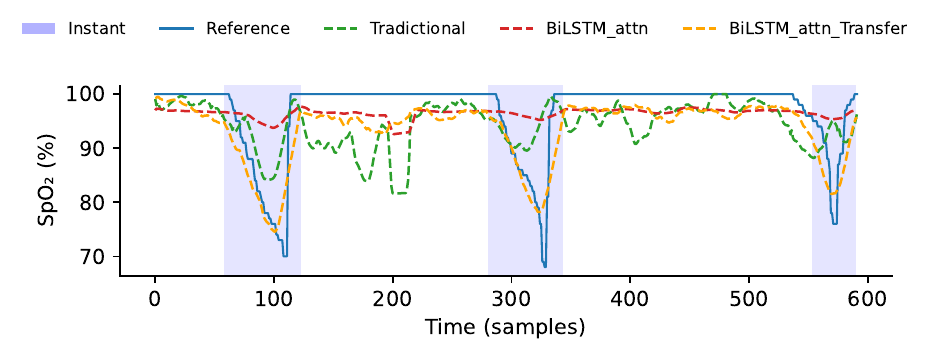}
    \caption{SpO$_2$ prediction results on a representative test case from private wearable dataset}
    \label{fig:webe_spo2}
\end{figure}

A significant oxygen desaturation event is typically defined as a $\geq$ 3\% drop in SpO$_2$ sustained for at least 10 seconds as measured by pulse oximetry \cite{kainulainen2020severe}. To further assess performance during instantaneous SpO$_2$ fluctuations, we define instant zones as time intervals where the total variation ($TV$) in SpO$_2$ exceeds 3\% within a sliding window of 10 seconds:

\begin{equation}
TV_i =  \sum_{j=i}^{i+10}  \left| y_{j+1} - y_j  \right|
\end{equation}

where $y_j$ represents the reference SpO$_2$ value at time step $j$. A time window is flagged as an instant zone if $TV_i$ $\geq$ 3\% at the starting index $i$. In these zones, we measured instantaneous errors MAE\textsubscript{ins} and RMSE\textsubscript{ins} to evaluate the model’s ability to track sharp desaturation or resaturation events. 

As shown in Table~\ref{tab:webe_spo2}, on the private dataset, both machine learning models outperformed the traditional calibration method. The BiLSTM+Attn+Transfer model achieved the best overall performance, with the lowest MAE of 2.624\% and MAE$_{\text{ins}}$ of 3.284\%. In addition, Fig.~\ref{fig:webe_spo2} illustrates a representative test case, highlighting the instant zones in purple. Compared to the reference SpO$_2$ (blue solid line), only the BiLSTM+Attn+Transfer model (orange dashed) closely follows the reference during rapid SpO$_2$ drops, effectively tracking transient desaturation..

% \begin{table}[htbp]
% \caption{SpO$_2$ Prediction Errors on the Case of Figure~\ref{fig:webe_spo2}}
% \begin{center}
% \begin{tabular}{|l|c|c|c|c|}
% \hline
% \textbf{Method} & \textbf{MAE} & \textbf{RMSE} & \textbf{MAE$_{\text{instant}}$} & \textbf{RMSE$_{\text{instant}}$} \\
% \hline
% Traditional          & 2.615 & 4.253 & 5.228 & 6.501 \\
% BiLSTM+Attn          & 6.688 & 7.003 & 8.266 & 8.553 \\
% BiLSTM+Attn+Transfer    & \textbf{1.292} & \textbf{1.735} & \textbf{2.048} & \textbf{2.436} \\
% \hline
% \end{tabular}
% \label{tab:case_spo2}
% \end{center}
% \end{table}

% For this specific case, the quantitative metrics in Table~\ref{tab:case_spo2} further validate our findings: BiLSTM+Attn+Transfer significantly outperforms all baselines, achieving the lowest MAE, RMSE, and instantaneous errors. This confirms that our method not only generalizes well through transfer learning, but also effectively captures transient oxygen desaturation events, which are critical for applications such as sleep apnea detection or ambulatory monitoring.

\section{Discussion} \label{Discussion}
Our results demonstrate the accurate SpO$_2$ estimation using PPG signals sampled at 25Hz. Despite the reduced temporal resolution, the BiLSTM+Attn model outperforms traditional calibration methods. While pretraining on the public dataset offers a strong initialization, the model struggles to generalize to wearable data due to domain shift and hardware differences. Transfer learning significantly improves performance in these non-clinical settings. Moreover, when fine-tuned on small amounts of wearable data, BiLSTM-based models with attention effectively capture rapid desaturation events, making them valuable for applications such as sleep apnea detection and remote monitoring. 

\section{Conclusion and Future Work}

We propose a transfer learning framework for rapid and accurate SpO$_2$ estimation on wearable devices using low-sampling-rate (25 Hz) PPG signals. By pretraining a BiLSTM with self-attention on a public clinical dataset and fine-tuning it on a small set of wearable data, our method eliminates the need for device-specific clinical calibration while substantially reducing power consumption. The proposed approach demonstrates strong performance on both public and private datasets, effectively capturing both average and instantaneous SpO$_2$ fluctuations and outperforming traditional calibration and non-transferred machine learning models. In future work, we will leverage temporal dependencies across sliding windows and employ better domain adaptation methods.

\bibliographystyle{ieeetr}
\bibliography{EMBC}

\end{document}